# Evaluating the Impact of Partial Volume Correction on FDG PET Radiomics Reproducibility in Lymphoma Lesions

Setareh Hasanabadi[1], Mohammad Saber Azimi[1], Mehrdad Bakhshayesh Karam[2], and Hossein Arabi[3]

[1] Department of Medical Radiation Engineering, Shahid Beheshti University, Tehran, Iran
[2] Masih Daneshvari Hospital, Tehran, Nuclear Medicine Department, Iran
[3] Division of Nuclear Medicine & Molecular Imaging, Geneva University Hospital, CH-1211, Geneva, Switzerland

## Abstract

**Purpose:** To evaluate how partial volume correction (PVC) affects the reproducibility of $^{18}$F-FDG PET radiomic features in lymphoma lesions, with respect to lesion volume and tissue type.
**Methods:** This single-center retrospective study included 131 newly diagnosed lymphoma patients (2014–2024) who underwent baseline $^{18}$F-FDG PET/CT. In total, 1,603 lesions (1,302 lymph nodes, 117 spleen/liver, 150 bone, and 34 bone/soft-tissue) were semi-automatically segmented and grouped by volume (<3, 3–10, 10–30, >30 mL) and tissue type. Ninety-three radiomic features were extracted from non-PVC and PVC images processed with the Richardson–Lucy (RL) and Reblurred Van Cittert (RVC) algorithms after isotropic resampling (3 mm) and discretization (0.25 SUV bin size), following IBSI guidelines. Reproducibility was quantified using the coefficient of variation (CoV) and the intraclass correlation coefficient ($ICC_2$, absolute agreement), with statistical comparisons performed via Mann–Whitney U tests and false-discovery-rate (FDR) correction.
**Results:** PVC significantly improved feature reproducibility, particularly for large lesions (>30 mL), with median $ICC_2 > 0.90$ across most feature categories (e.g., First-Order = 0.99, GLSZM = 0.97, NGTDM = 0.97). Small lesions (<3 mL) showed lower stability ($ICC_2$ = 0.84–0.94) and higher CoV (0.09–0.21), mainly in texture-based features. First-Order and GLCM features were the most robust overall ($ICC_2$ = 0.92–0.99; CoV = 0.07–0.11). Bone and spleen lesions exhibited the highest reproducibility (median $ICC_2 \approx 0.95$), whereas lymph-node and liver features were more variable. All volume- and tissue-dependent differences remained significant after FDR correction ($p < 0.05$).
**Conclusion:** PVC using RL and RVC markedly enhances FDG-PET radiomic reproducibility in lymphoma, particularly for larger and structurally uniform lesions. Robust features such as First-Order and GLCM can support standardized radiomics workflows and the development of reliable biomarkers for prognosis and personalized therapy. Multicenter validation is warranted to confirm generalizability beyond a single-center setting.

**Keywords:** $^{18}$F-FDG PET, lymphoma, radiomics, reproducibility, partial volume correction (PVC), lesion volume.

## 1. Introduction

$^{18}$F-Fluorodeoxyglucose positron emission tomography ($^{18}$F-FDG PET) plays a central role in lymphoma management, supporting diagnosis, staging, treatment planning, and therapy response assessment (1-3). By quantifying tumor metabolism, $^{18}$F-FDG PET provides valuable insights into disease burden and heterogeneity, which are essential for personalized treatment strategies in both Hodgkin and non-Hodgkin lymphoma (4).

Radiomics, the high-throughput extraction of quantitative features from medical images, has expanded the diagnostic and prognostic capabilities of PET. Radiomic features describing tumor intensity, texture, and shape can predict outcomes such as progression-free and overall survival, often outperforming traditional indices like the International Prognostic Index (IPI) (5-9). However, the clinical translation of radiomics remains limited by poor feature reproducibility, which undermines the stability of predictive models across scanners and imaging centers.

One major source of variability arises from partial volume effects (PVE), caused by the limited spatial resolution of PET systems (typically 4–6 mm full-width at half-maximum). PVE leads to signal spill-over and underestimation of tracer uptake, particularly in small (< 3 mL) or heterogeneous lesions (10, 11). These artifacts distort both intensity- and texture-based metrics, compromising quantitative reliability.

Partial volume correction (PVC) techniques, including iterative methods such as Richardson–Lucy (RL) and Reblurred Van Cittert (RVC), have been widely adopted to mitigate PVE and improve quantitative accuracy (12-14). PVC has demonstrated benefits in solid tumors like non-small-cell lung cancer (NSCLC), improving the reproducibility of texture and metabolic parameters, especially in larger lesions (15). However, its impact on radiomic stability in lymphoma, where both nodal and extranodal sites exhibit high biological and anatomical heterogeneity, remains largely unexplored.

Lymphoma presents unique challenges for radiomics due to its multi-organ involvement and broad lesion-size distribution. Nodal and extranodal lesions (e.g., in spleen, liver, and bone) exhibit distinct metabolic and textural profiles that may differently influence radiomic reproducibility (16). While previous studies have looked at radiomic reproducibility in brain or lung imaging, the combination of lesion volume and tissue type in lymphoma has not been systematically studied (17). Moreover, the effect of PVC across different lymphoma sites remains largely unexplored.

The present study provides the first comprehensive evaluation of $^{18}$F-FDG PET radiomic reproducibility following PVC in lymphoma, encompassing both nodal and extranodal lesions. By quantifying how PVC (via RL and RVC) affects feature stability across lesion sizes and tissue categories, this work delivers crucial evidence for developing standardized radiomic pipelines. The findings aim to identify robust and generalizable features suitable for multicenter studies, advancing reproducible imaging biomarkers for personalized therapy in lymphoma.

## 2. Material and Method

### 2.1 Data acquisition

This study included patients who underwent $^{18}$F-FDG PET/CT at Masih Daneshvari Hospital (Tehran, Iran) between 2014 and 2024. The study protocol was approved by the Medical Ethics Committee of Shahid Beheshti University of Medical Sciences (approval code: IR.SBMU.NRITLD.REC.1402.060). Because this was a non-interventional imaging analysis using anonymized data, the requirement for informed consent was waived.

Eligible participants were newly diagnosed lymphoma patients who underwent baseline $^{18}$F-FDG PET/CT for staging and whose diagnosis was confirmed by histopathology. Exclusion criteria were as follows: (1) non-original or incomplete PET/CT datasets, (2) negative or indeterminate findings, (3) suspected infections or inflammatory conditions, (4) known hepatic fibrosis or cirrhosis affecting physiological liver uptake, (5) concurrent or recent malignancies (e.g., breast cancer), and (6) scans reconstructed using non-standard protocols.

### 2.2 Imaging protocol

PET/CT imaging was performed on a GE Discovery 690 scanner (GE Healthcare, Milwaukee, WI, USA) equipped with time-of-flight (TOF) capability and a 64-slice CT system. Whole-body acquisitions were obtained from the vertex to mid-thigh. Image reconstruction was performed using the Vue Point HD Sharp (VPHDS) iterative algorithm provided by the manufacturer.

All patients fasted for at least 6 hours before tracer administration, ensuring a blood glucose level <140 mg/dL at the time of injection. A mean interval of 60 ± 10 minutes (range, 45–75 minutes) between tracer injection and image acquisition was maintained. Each PET bed position was acquired for 2–3 minutes. The PET slice thickness was 3.75 mm, while low-dose CT was acquired with 1.33–2.5 mm slice thickness. CT parameters included a tube voltage of 120 kVp, tube current modulation between 50–150 mA (automatically adjusted to patient body habitus), and a helical pitch of 0.9. PET data were corrected for scatter, randoms, attenuation (using CT-based attenuation correction), and decay before reconstruction.

.

**2.3 Image evaluation and lesion segmentation**

All PET/CT datasets were reviewed by a board-certified nuclear medicine physician with over ten years of experience and a board-certified radiologist with more than 34 years of experience. The primary evaluation involved disease staging and identification of both nodal and extranodal disease sites. For each patient, all metabolically active lesions were considered, including both nodal and extranodal sites (e.g., spleen, liver, bone, and soft tissue), with no numerical limit on the number of lesions per patient (Figure 1).

Lesion segmentation was performed using a semi-automated, gradient-based method (18), implemented as an extension within the 3D Slicer platform (version 5.6.0) (19). The nuclear medicine physician visually inspected and manually refined each delineation to ensure anatomical and metabolic accuracy. A second reader independently verified a subset of the segmentations for quality control and reproducibility.

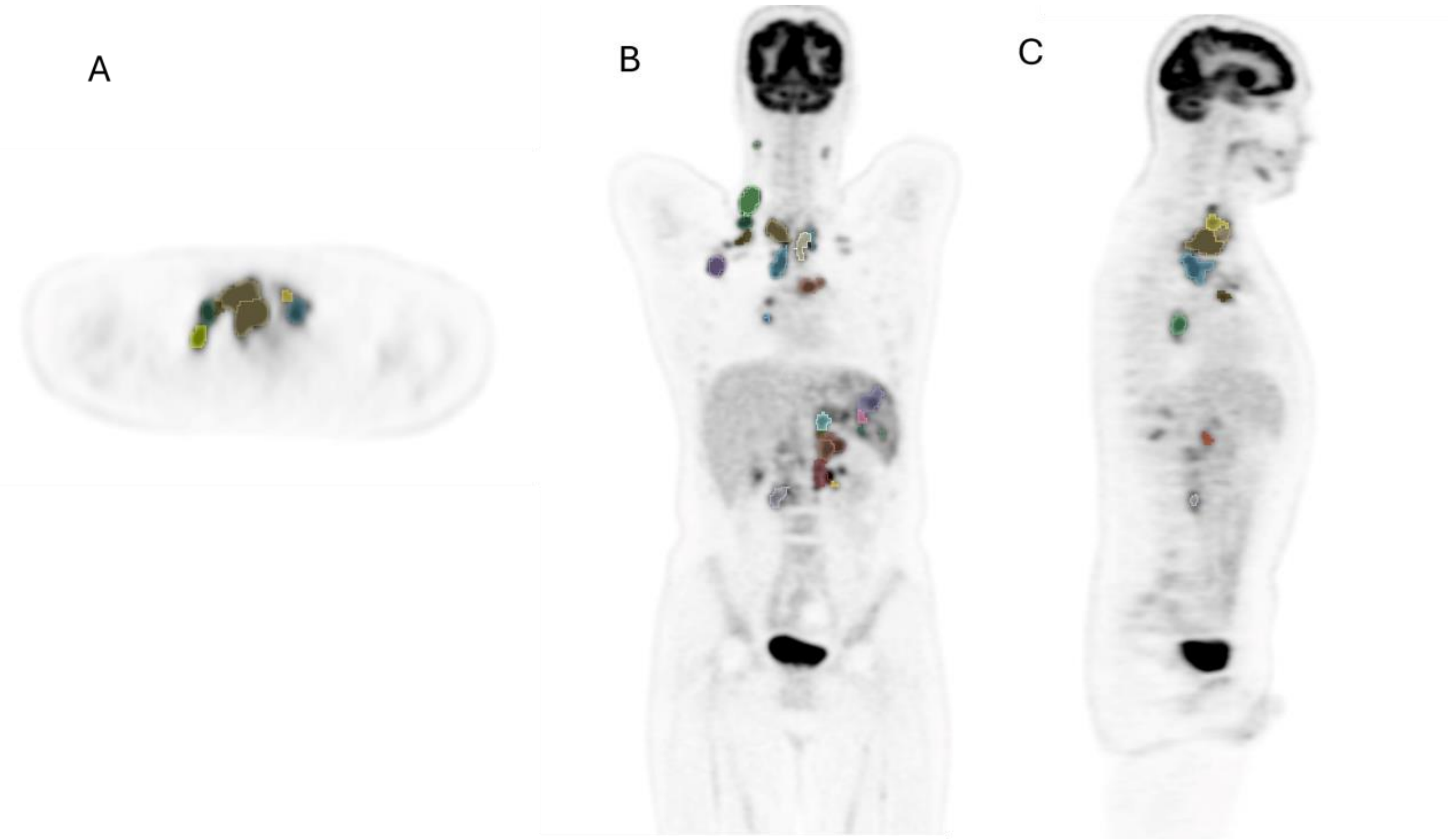

**Figure 1.** Representative example of the semi-automatic graph-based segmentation method (A: transaxial, B: coronal, and C: sagittal view).

## 2.4 Image preprocessing and feature extraction

PET images were resampled to an isotropic voxel size of 3 × 3 × 3 mm³ to ensure rotational and inter-scanner consistency. All images were converted into standardized uptake value (SUV) maps and discretized using a fixed bin width of 0.25 SUV, following the Image Biomarker Standardization Initiative (IBSI) guidelines (20) . Radiomic features were extracted using the PyRadiomics package (version 3.0.1) (21) integrated within 3D Slicer (19). Feature extraction was performed separately for each lesion on both non–PVC and PVC-corrected PET images (processed with the RL and RVC algorithms). A total of 93 radiomic features were computed and categorized into six groups:

- First-Order Statistics (18 features)
- Gray-Level Co-occurrence Matrix (GLCM, 24 features)

- Gray-Level Run Length Matrix (GLRLM, 16 features)
- Gray-Level Dependence Matrix (GLDM, 14 features)
- Gray-Level Size Zone Matrix (GLSZM, 16 features)
- Neighborhood Gray Tone Difference Matrix (NGTDM, 5 features)

All features were extracted in accordance with IBSI definitions to ensure cross-platform comparability.

### 2.5 Partial Volume Correction (PVC)

Partial volume correction was performed using the PETPVC toolbox (version 1.2.0; University College London, London, UK) (12), a C++-based software package built on the Insight Segmentation and Registration Toolkit (ITK) framework. This toolbox implements several established post-reconstruction PVC algorithms. In this study, two widely used deconvolution-based methods, the Richardson–Lucy (RL) and Reblurred Van Cittert (RVC) algorithms, were applied to all reconstructed PET images.

Both RL and RVC are voxel-wise, mask-independent iterative deconvolution approaches that use knowledge of the imaging system's point spread function (PSF) to iteratively compensate for partial volume effects (PVE). The algorithms operate in the image domain and do not require anatomical segmentation masks. Each iteration estimates a corrected image by convolving the previous estimate with a 3D Gaussian kernel representing the PSF, followed by a normalization step to prevent noise amplification. In this study, 10 iterations were applied for RL and 8 iterations for RVC, as recommended by prior validation work (30).

For the Discovery 690 PET/CT scanner (GE Healthcare, Milwaukee, WI, USA) used in this study, equipped with time-of-flight (TOF) and PSF modeling via Vue Point HD Sharp (VPHDS) reconstruction, the intrinsic system spatial resolution was characterized by a three-dimensional Gaussian PSF with a full width at half maximum (FWHM) of approximately 4.3 mm in the transaxial (x, y) directions and 4.8–5.0 mm in the axial (z) direction (22). These parameters were incorporated into PETPVC as the Gaussian kernel under a shift-invariant assumption for both RL and RVC deconvolution procedures.

**2.6 Statistical analysis**

Radiomic features were extracted from all segmented lesions across nodal (lymph node) and extranodal regions (spleen, liver, bone, and soft tissue). Lesions were further stratified into four volume groups: <3 mL, 3–10 mL, 10–30 mL, and >30 mL. Shape-based descriptors (e.g., mesh volume, surface area) were excluded from reproducibility analysis to focus on intensity- and texture-derived features. Non-numeric or missing values were removed prior to computation.

Two complementary statistical metrics were used to evaluate radiomic feature reproducibility across imaging methods:

1. Coefficient of Variation (CoV):

   The CoV was computed for each feature within individual lesions to quantify intra-lesion variability. For each feature class, the median of CoV values were reported.

2. Intraclass Correlation Coefficient (ICC):

   Reproducibility between imaging methods (non-PVC, RL, and RVC) was quantified using ICC(2,1) for absolute agreement for consistency, both derived from a two-way mixed-effects ANOVA model. Lesion identifiers were treated as random effects and imaging methods as fixed raters. Analytical 95% confidence intervals (CIs) were estimated using the F distribution. Following established conventions, ICC values were interpreted as: poor (<0.5), moderate (0.5–0.75), good (0.75–0.9), and excellent (>0.9) agreement.

Differences across imaging methods were examined using the Friedman test, with subsequent linear mixed-effects modeling to account for within-patient correlations (patient ID as a random effect). Multiple testing corrections were performed using the Benjamini–Hochberg false discovery rate (FDR) procedure.

For comparisons across lesion volumes and tissue categories, pairwise Mann–Whitney U tests were applied to ICC(2,1) distributions. The ΔMedian ICC was defined as the difference between the median ICCs of two groups, and all p-values were adjusted for multiple comparisons using FDR correction.

All data management and statistical analyses were performed in Python (v3.10) using the libraries Pingouin (v0.5.5), SciPy (v1.16.2), and StatsModels (v0.14.5), with numerical computations handled via NumPy (v2.0.2).

## 3. Results

The study included 131 patients with a mean age of 40.8 ± 20.4 years, 70 (53.4%) male and 61 (46.6%) female. Regarding disease type, 73 (58.9%) had HL and 51 (41.1%) had NHL. A total of 1,603 lesions were analyzed, and 93 radiomic features were successfully extracted from both non-PVC and PVC-corrected PET images. Reproducibility was quantified using the intraclass correlation coefficient ($ICC_2$) for absolute agreement and the coefficient of variation (CoV) for intra-lesion variability.

### 3.1 Effect of lesion volume on feature reproducibility

Figure 2A summarizes $ICC_2$ across four lesion-volume groups (< 3 mL, 3–10 mL, 10–30 mL, and > 30 mL). Feature reproducibility increased consistently with lesion size. Small lesions (< 3 mL) exhibited the lowest stability (median $ICC_2$ = 0.84–0.94), whereas large lesions (> 30 mL) achieved excellent reproducibility (median $ICC_2 > 0.95$). Correspondingly, CoV values decreased with increasing volume (Figure 2B), confirming reduced variability in larger lesions. First-Order and GLCM features were the most robust (CoV ≈ 0.04–0.10), while GLSZM and NGTDM features were highly variable in small lesions (CoV ≈ 0.17–0.21). Table 1 presents the pairwise Mann–Whitney U comparisons between lesion-volume groups. Lesions < 3 mL were significantly less reproducible than all larger groups (ΔMedian $ICC_2$ = −0.035 to −0.060, all FDR < 0.001), confirming that reproducibility improves with increasing lesion size. Intermediate volumes (3–10 mL vs. 10–30 mL) also showed significant yet smaller differences.

### 3.2 Effect of tissue type on reproducibility

Figure 3A and 2B display $ICC_2$ and CoV distributions across five tissue categories (lymph node, spleen, liver, bone, and bone/soft tissue). Bone lesions demonstrated the highest reproducibility (median $ICC_2 \approx 0.95$) with low CoV ($\approx 0.08$), whereas lymph-node and spleen lesions showed moderate stability ($ICC_2 = 0.88–0.94$). Liver and mixed bone–soft-tissue regions exhibited greater variability, particularly in texture-related features. Table 2 demonstrates that lymph-node features were significantly less reproducible than spleen ($\Delta$Median $ICC_2 = -0.015$, FDR = 0.0263) but slightly more stable than bone ($\Delta$Median $ICC_2 = 0.020$, FDR = 0.0246). No significant differences were observed for liver or mixed tissues after FDR correction.

### 3.3 Feature-class–specific reproducibility

Table 3 summarizes $ICC_2$ and CoV values across six feature classes. First-Order and GLCM features exhibited the highest robustness ($ICC_2 = 0.92–0.99$, CoV = 0.04–0.11) across all lesion sizes and tissues. GLDM and GLRLM features showed moderate reproducibility ($ICC_2 = 0.88–0.97$) that improved with lesion size. Conversely, GLSZM and NGTDM features were highly sensitive to both lesion volume and tissue heterogeneity, showing poor stability in small or heterogeneous regions but reaching excellent agreement ($ICC_2 \approx 0.96$) in larger, homogeneous lesions.

### 3.4 Heatmap-based visualization of reproducibility patterns

Figure 4 visualizes radiomic reproducibility significance as $-\log_{10}(ICC_2$ p-values). Features in larger lesions and bone tissues exhibit darker shades, indicating higher statistical significance. Figure 5 presents categorical $ICC_2$ reproducibility maps across lesion volumes and tissues, showing that small lesions and heterogeneous regions (e.g., lymph node, spleen) have a higher proportion of moderate or poor categories, while larger and homogeneous lesions (e.g., bone) predominantly fall within the excellent range. Figure 6 shows categorical CoV heatmaps, where higher reproducibility corresponds to lower CoV values (green shades). Variability increases notably in small lesions (<3 mL) and complex tissues (liver, lymph node), especially in GLSZM and NGTDM feature classes.

### 3.5 Summary of trends

Figures 2–6 collectively demonstrate that both lesion volume and tissue type strongly influence radiomic reproducibility. Small lesions and metabolically heterogeneous tissues (e.g., lymph node, spleen) show reduced stability, whereas larger and structurally uniform lesions (e.g., bone) exhibit consistent, high reproducibility. Among all feature families, First-Order and GLCM metrics remain the most reliable, while GLSZM and NGTDM are highly sensitive to lesion size and tissue heterogeneity.

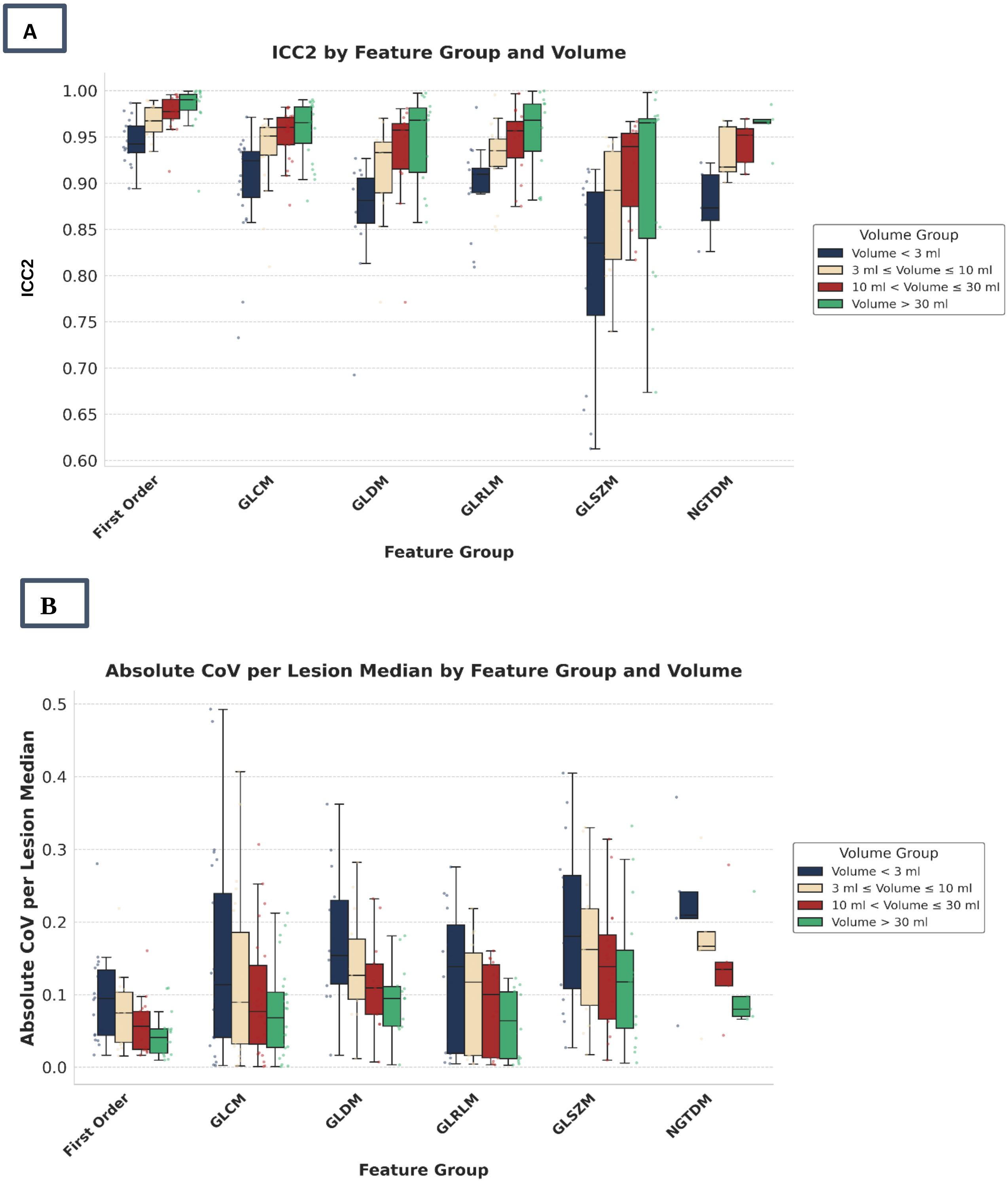

**Figure 2.** Radiomic feature reproducibility across lesion volume groups. **Panel A**: ICC2 (absolute agreement) per feature group (First Order, GLCM, GLDM, GLRLM, GLSZM, NGTDM) for lesions of <3 ml (dark blue), 3–10 ml (wheat), 10–30 ml (red), and >30 ml (green). **Panel B:** Absolute median CoV per lesion for the same feature groups and volume categories. Boxplots show distributions, overlaid points represent individual values. Higher ICC2 and lower CoV indicate greater reproducibility.

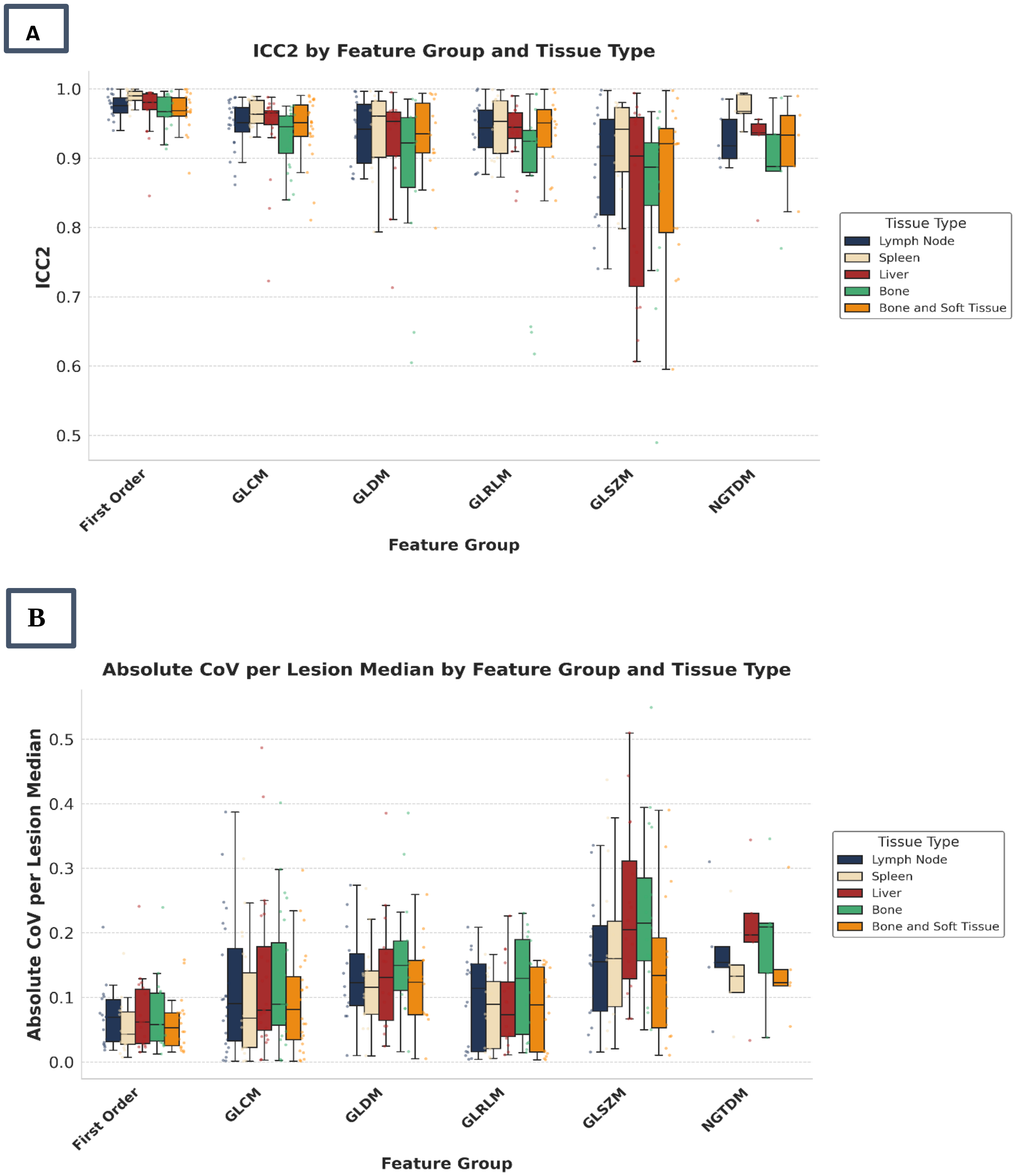

**Figure 3.** Radiomic feature reproducibility across tissue types. **Panel A**: ICC2 (absolute agreement) per feature group (First Order, GLCM, GLDM, GLRLM, GLSZM, NGTDM) for different tissues: Lymph Node (Nodal), Spleen (Extranodal), Liver (Extranodal), Bone (Extranodal), Bone and Soft Tissue (Extranodal). **Panel B**: Absolute median CoV per lesion for the same feature groups and tissue types. Boxplots show distributions, overlaid points represent individual values. Higher ICC2 and lower CoV indicate greater reproducibility.

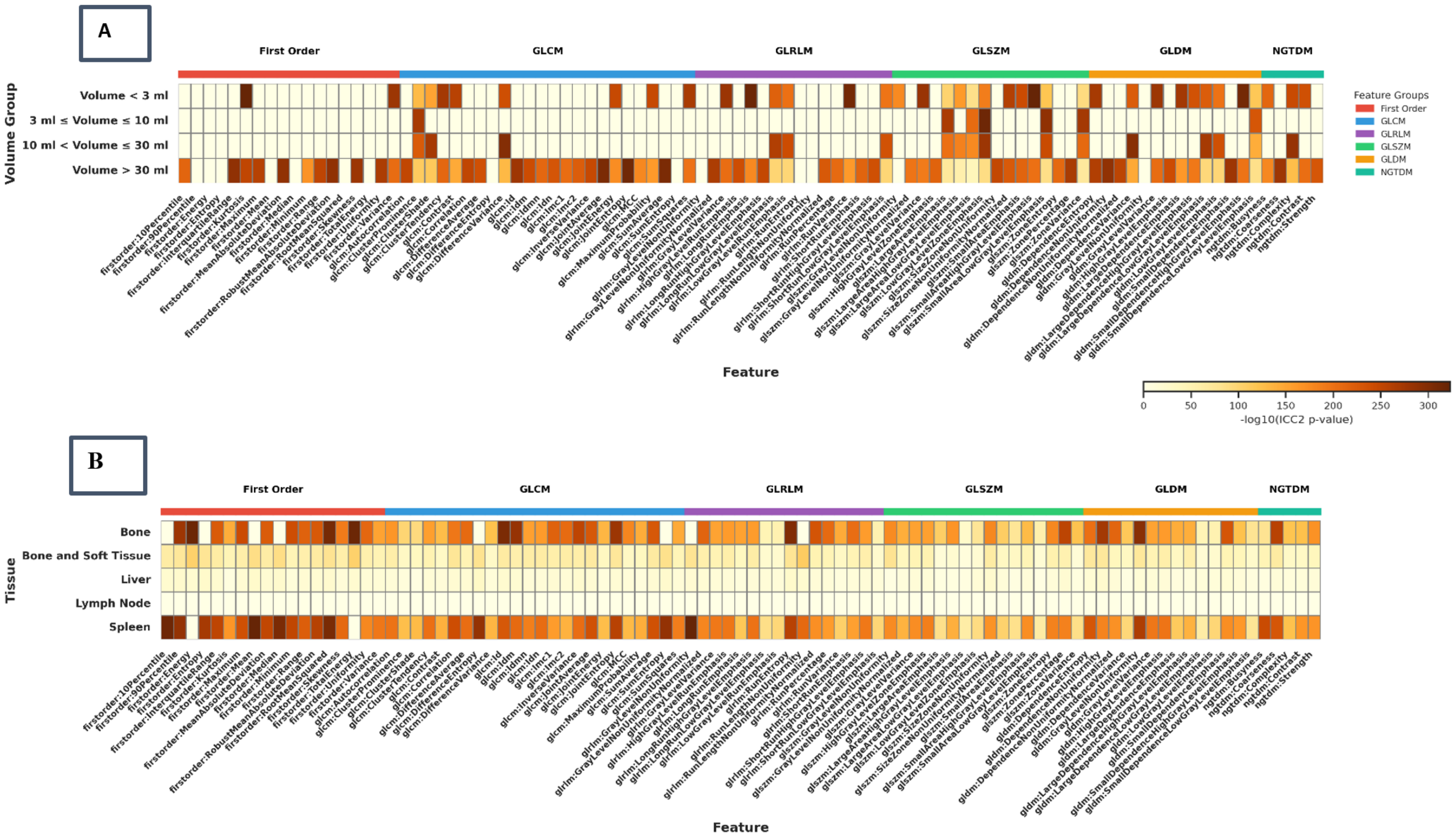

**Figure 4.** Radiomic feature reproducibility visualized as heatmaps based on $-\log_{10}$(ICC$_2$ p-values). **Panel A:** Statistical significance ($-\log_{10}$(p)) of ICC$_2$ values across lesion volume groups (< 3 mL, 3–10 mL, 10–30 mL, and > 30 mL). **Panel B:** Statistical significance ($-\log_{10}$(p)) of ICC$_2$ values across tissue categories (bone, bone and soft tissue, liver, lymph node, and spleen). Columns represent individual radiomic features grouped by category (First Order, GLCM, GLRLM, GLSZM, GLDM, NGTDM). Higher $-\log_{10}$(p) values (darker shades) indicate features with more statistically significant ICC$_2$ reproducibility.

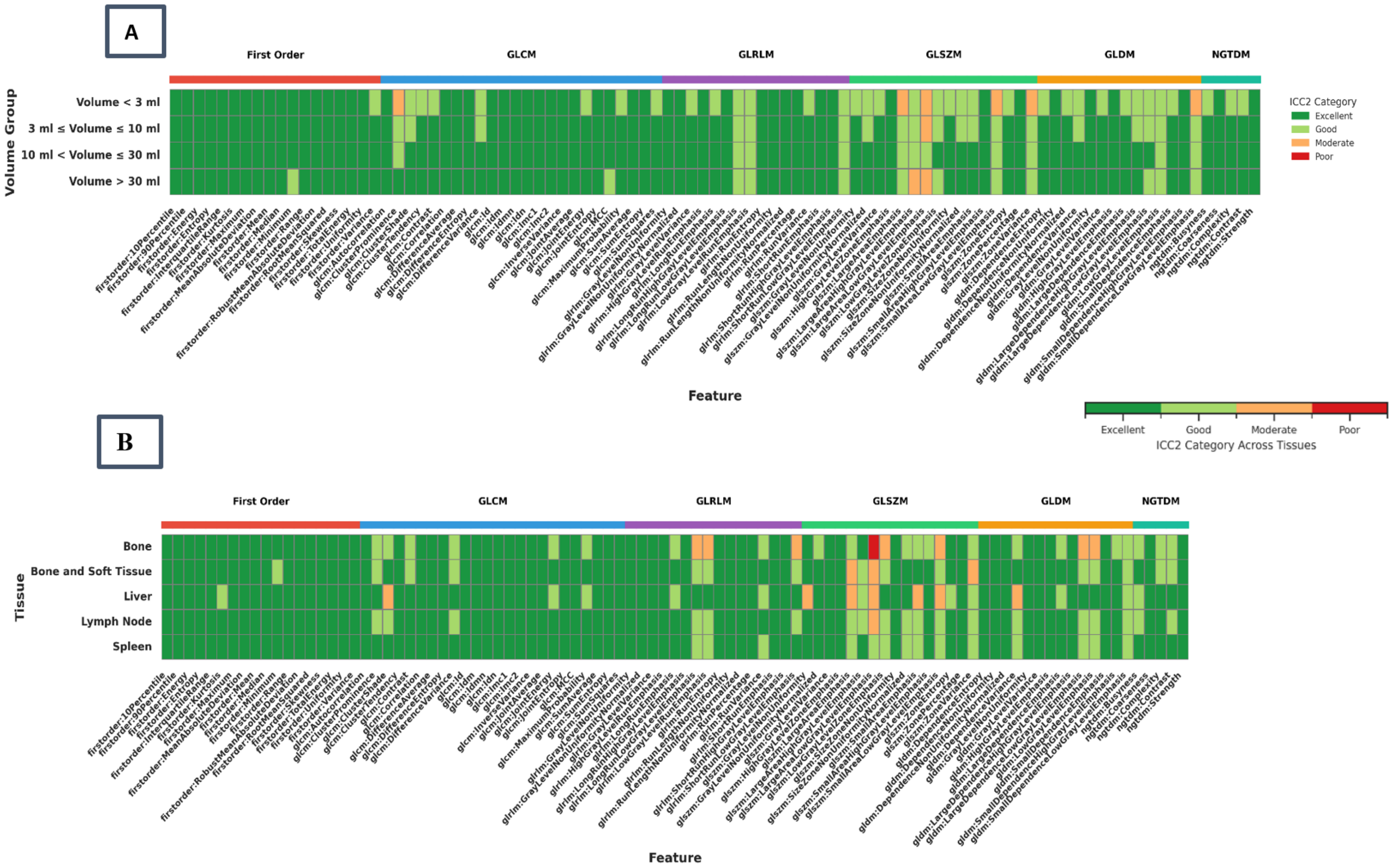

**Figure 5.** Categorical heatmap of radiomic feature reproducibility ($ICC_2$ categories) across lesion volume and tissue groups. Each cell represents the qualitative category of $ICC_2$ reproducibility: Excellent ($\geq 0.90$), Good (0.75–0.90), Moderate (0.50–0.75), and Poor ($< 0.50$). Panel A: $ICC_2$ categories across four lesion volume groups ($< 3$ mL, 3–10 mL, 10–30 mL, and $> 30$ mL), illustrating the dependence of feature reproducibility on lesion size. Panel B: $ICC_2$ categories across tissue types (lymph node, spleen, liver, bone, and bone/soft tissue), showing inter-tissue differences in feature stability. Color bars indicate reproducibility categories from Excellent (green) to Poor (red).

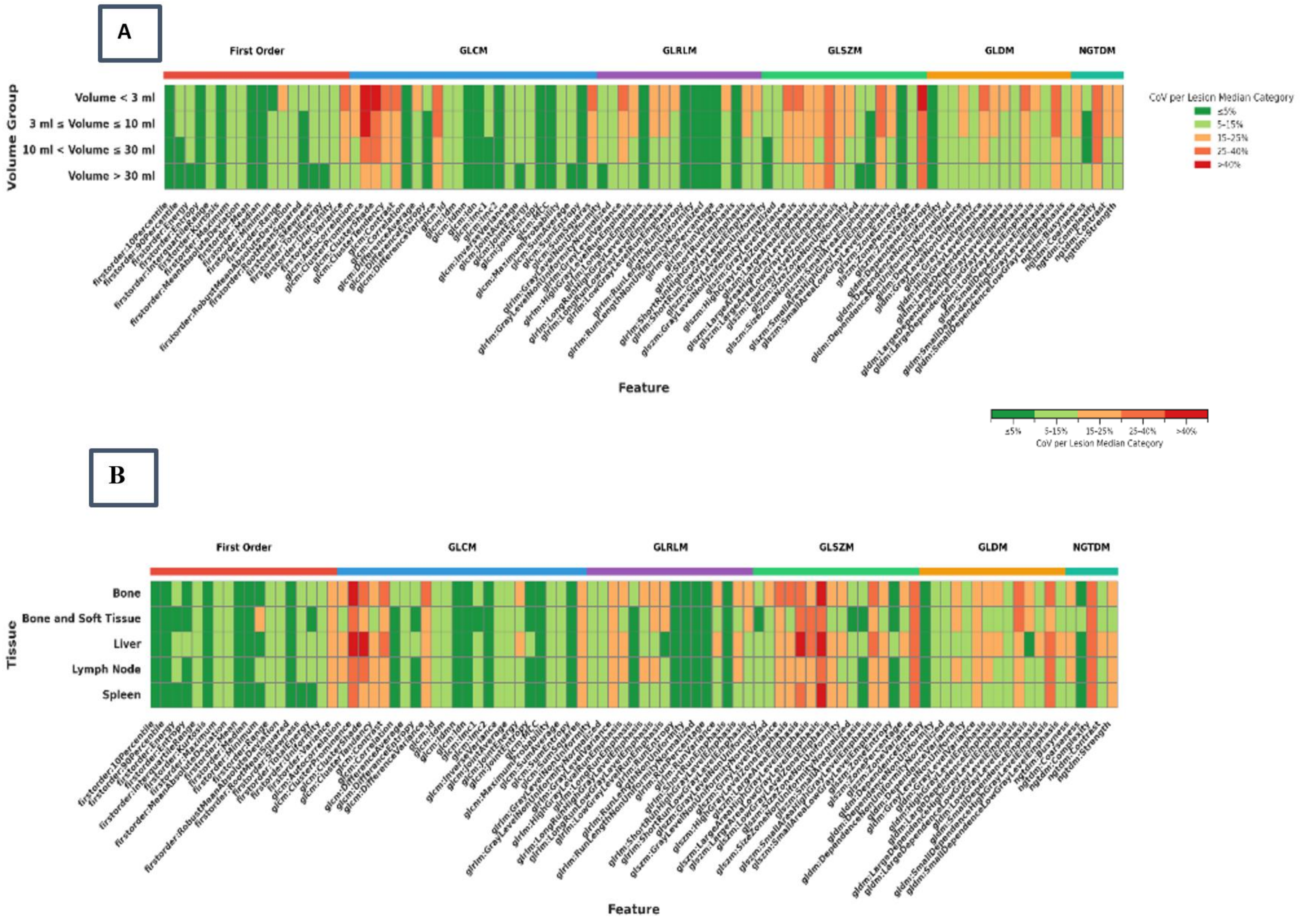

**Figure 6.** Categorical heatmap of absolute Coefficient of Variation (CoV) across lesion volume and tissue groups. Each cell represents the median absolute CoV category for a given radiomic feature: ≤ 5% (Excellent), 5–15% (Good), 15–25% (Moderate), 25–40% (High), and > 40% (Very High). Panel A: CoV categories across four lesion-volume groups (< 3 mL, 3–10 mL, 10–30 mL, > 30 mL), illustrating the effect of lesion size on measurement variability. Panel B: CoV categories across five tissue types (lymph node, spleen, liver, bone, and bone/soft tissue), demonstrating inter-tissue differences in radiomic feature stability. Color scales indicate relative variability, with greener shades corresponding to lower variability (higher reproducibility) and redder shades to higher variability (lower reproducibility).

**Table 1.** Pairwise comparison of lesion-volume groups based on $ICC_2$ (Mann–Whitney U test). ΔMedian $ICC_2$ represents the difference in median $ICC_2$ values between groups. FDR-adjusted p-values are reported; significant results (FDR < 0.05) are indicated. Smaller lesions (< 3 mL) show significantly lower reproducibility than larger ones across all comparisons.

| **Comparison** | **ΔMedian ICC2** | **p-value (Mann-Whitney)** | **FDR** | **Interpretation** |
|---|---|---|---|---|
| Volume < 3 ml vs 3 ml ≤ Volume ≤ 10 ml | -0.035 | 6.03E-07 | 0 | Significant |
| Volume < 3 ml vs 10 ml < Volume ≤ 30 ml | -0.049 | 1.31E-12 | 0 | Significant |
| Volume < 3 ml vs Volume > 30 ml | -0.06 | 5.12E-14 | 0 | Significant |
| 3 ml ≤ Volume ≤ 10 ml vs 10 ml < Volume ≤ 30 ml | -0.014 | 0.00089 | 0.0011 | Significant |
| 3 ml ≤ Volume ≤ 10 ml vs Volume > 30 ml | -0.024 | 2.18E-07 | 0 | Significant |
| 10 ml < Volume ≤ 30 ml vs Volume > 30 ml | -0.011 | 0.00241 | 0.0024 | Significant |

**Table 2.** Pairwise comparison of tissue types based on $ICC_2$ (Mann–Whitney U test). ΔMedian $ICC_2$ represents the difference in median $ICC_2$ values between tissue categories. FDR-adjusted p-values are reported; results with FDR < 0.05 are considered significant ("NS" = not significant). Lymph-node features were slightly less reproducible than spleen but more stable than bone; bone and spleen generally exhibited the highest reproducibility.

| **Comparison** | **ΔMedian ICC2** | **p-value (Mann-Whitney)** | **FDR** | **Interpretation** |
|---|---|---|---|---|
| Lymph Node vs Spleen | -0.015 | 0.015807 | 0.0263 | Significant |
| Lymph Node vs Liver | -0.003 | 0.709021 | 0.7728 | NS |
| Lymph Node vs Bone | 0.02 | 0.010454 | 0.0246 | Significant |
| Lymph Node vs Bone and Soft Tissue | 0.003 | 0.591542 | 0.7394 | NS |
| Spleen vs Liver | 0.011 | 0.003721 | 0.0186 | Significant |
| Spleen vs Bone | 0.035 | 1.24E-06 | 0 | Significant |
| Spleen vs Bone and Soft Tissue | 0.018 | 0.006242 | 0.0208 | Significant |
| Liver vs Bone | 0.023 | 0.012306 | 0.0246 | Significant |
| Liver vs Bone and Soft Tissue | 0.007 | 0.772787 | 0.7728 | NS |
| Bone vs Bone and Soft Tissue | -0.017 | 0.037932 | 0.0542 | NS |

**Table 3.** Median Coefficient of Variation (CoV) and Intraclass Correlation Coefficient ($ICC_2$) of radiomic features across lesion-volume ranges. Median CoV and $ICC_2$ values are reported for six feature categories (First Order, GLCM, GLDM, GLRLM, GLSZM, NGTDM) across four lesion-volume groups (< 3 mL, 3–10 mL, 10–30 mL, > 30 mL). "n" denotes the number of lesions in each category. Reproducibility (higher $ICC_2$ / lower CoV) improves consistently with increasing lesion volume across all feature classes.

| Feature Category | Volume Category | Number of Lesions (n) | Median CoV | Median ICC |
|---|---|---|---|---|
| First order | Volume < 3 ml | 337 | 0.094 | 0.942 |
| First order | 3 ml ≤ Volume ≤ 10 ml | 516 | 0.075 | 0.967 |
| First order | 10 ml < Volume ≤ 30 ml | 315 | 0.056 | 0.978 |
| First order | Volume > 30 ml | 134 | 0.041 | 0.990 |
| GLCM | Volume < 3 ml | 337 | 0.114 | 0.924 |
| GLCM | 3 ml ≤ Volume ≤ 10 ml | 516 | 0.090 | 0.951 |
| GLCM | 10 ml < Volume ≤ 30 ml | 315 | 0.077 | 0.961 |
| GLCM | Volume > 30 ml | 134 | 0.068 | 0.965 |
| GLDM | Volume < 3 ml | 337 | 0.154 | 0.881 |
| GLDM | 3 ml ≤ Volume ≤ 10 ml | 516 | 0.127 | 0.933 |
| GLDM | 10 ml < Volume ≤ 30 ml | 315 | 0.109 | 0.958 |
| GLDM | Volume > 30 ml | 134 | 0.095 | 0.968 |
| GLRM | Volume < 3 ml | 337 | 0.139 | 0.910 |
| GLRM | 3 ml ≤ Volume ≤ 10 ml | 516 | 0.117 | 0.935 |
| GLRM | 10 ml < Volume ≤ 30 ml | 315 | 0.100 | 0.957 |
| GLRM | Volume > 30 ml | 134 | 0.064 | 0.968 |
| GLSZM | Volume < 3 ml | 337 | 0.175 | 0.835 |
| GLSZM | 3 ml ≤ Volume ≤ 10 ml | 516 | 0.162 | 0.892 |
| GLSZM | 10 ml < Volume ≤ 30 ml | 315 | 0.139 | 0.940 |
| GLSZM | Volume > 30 ml | 134 | 0.117 | 0.965 |
| NGTDM | Volume < 3 ml | 337 | 0.209 | 0.873 |
| NGTDM | 3 ml ≤ Volume ≤ 10 ml | 516 | 0.166 | 0.917 |
| NGTDM | 10 ml < Volume ≤ 30 ml | 315 | 0.135 | 0.952 |
| NGTDM | Volume > 30 ml | 134 | 0.080 | 0.966 |

## 4. Discussion

This study shows that PVC using deconvolution-based algorithms, Richardson–Lucy (RL) and Reblurred Van Cittert (RVC), substantially improves the reproducibility of $^{18}$F-FDG PET radiomic features in lymphoma across nodal and extranodal sites. Consistent with the hypothesis that PVC mitigates PVE, reproducibility increased with lesion size and varied by tissue type. These findings are supported by the convergent patterns observed in Figures 2–6 and the statistics summarized in Tables 1–3.

The present study demonstrates that lesion size plays a decisive role in determining radiomic feature reproducibility after PVC. Larger lesions (>30 ml) consistently exhibited excellent $ICC_2$ values (0.965–0.990), whereas smaller ones (<3 ml) showed moderate-to-high reproducibility (0.835–0.942). This size-dependent pattern aligns with the known physical limitations of PET imaging, where PVE dominate in small objects because of the system's finite spatial resolution (4.3–5.0 mm FWHM in the GE Discovery 690 used here) (23). By compensating for PVE, PVC effectively restores spatial contrast and stabilizes textural metrics (24). The significant $ICC_2$ differences across lesion-size groups ($\Delta$Median $ICC_2 = -0.035$ to $-0.060$, FDR < 0.001; Table 1) thus reaffirm that lesion volume remains a primary determinant of radiomic stability even after correction.

While the improvement in reproducibility is evident, the uniformly high $ICC_2$ values (>0.90 across most categories; Table 3) suggest that methodological factors may also contribute. Features with inherently low sensitivity to intensity variations, such as First Order statistics (e.g., firstorder:10Percentile), achieved $ICC_2$ up to 0.990 with CoV as low as 0.041, whereas complex textural matrices (GLSZM, NGTDM) exhibited reduced stability in smaller lesions but still reached $ICC_2 \geq 0.965$ for large volumes (Table 3, Figure 6). This trend is broadly consistent with phantom studies showing that PVC enhances reproducibility mainly in volumes above 10 ml, often achieving ICC ≈ 0.85 for First Order and GLCM features but rarely exceeding 0.95 (25). Likewise, in brain PET research, RL and RVC achieved ICC ≥ 0.75 for most features, with only highly homogeneous regions such as the cerebellum approaching ICC ≥ 0.9 (14). Hence, the near-universal $ICC_2 > 0.95$ observed in our large lesions likely reflects both effective PVE correction and the high standardization of our imaging workflow (single scanner, VPHDS reconstruction,

IBSI-compliant discretization (25)), yet could also indicate a ceiling effect or limited variability between the RL and RVC algorithms.

When contextualized with previous reports, these findings highlight both progress and remaining limitations. In non-small cell lung cancer (NSCLC), PVC improved reproducibility for lesions >3 ml but with lower median ICC (0.823) and higher CoV (30.37%) compared to our results (15). This difference may stem from respiratory motion artifacts in thoracic PET, which exacerbate PVE and degrade texture fidelity. By contrast, in our lymphoma cohort, where lesions are distributed systemically and motion is less pronounced, PVC yielded stronger gains, particularly for larger lesions (Table 3, Figure 6). Nonetheless, even though GLSZM and NGTDM remained sensitive to small volumes (CoV 0.175–0.209), their improved stability in larger lesions aligns with previous phantom data showing up to 30% variability for zone-based features in small volumes (~2.5 ml) (25). Notably, the nearly uniform $ICC_2$ >0.90 in our dataset surpasses values typically reported in the literature (often ≤0.95 even after PVC (26)), which may reflect reduced segmentation variability from expert-supervised semi-automatic delineation (19) and the controlled single-scanner setup (8, 10).

Tissue-specific analyses further demonstrate how biological heterogeneity affects feature robustness. Bone and spleen lesions displayed the highest reproducibility (median $ICC_2 \approx 0.95$), while lymph nodes and liver showed lower stability (0.88–0.95; CoV 10.3–13.4%) (Figure 5, Table 2). This gradient mirrors previous evidence that homogeneous tissues, such as bone or brain white matter, yield higher reproducibility than heterogeneous, metabolically variable organs like the liver (17). Within lymphoma, this distinction is clinically meaningful: extranodal sites like spleen and liver often exhibit diffuse uptake patterns, which likely explain the lower stability of GLDM and GLRLM features observed in these organs (27). High $ICC_2$ values may partly reflect methodological factors, as single-center design and similar RL/RVC algorithms can inflate reproducibility (19, 28) . Broader validation with non-PVC and deep-learning PVC methods (11) is needed to confirm robustness. In lymphoma, baseline $^{18}$F-FDG PET radiomics predicts outcomes (6, 29); our results show PVC improves feature stability, especially in large lesions where PVE, not motion as in NSCLC (15), is dominant.

While our results demonstrate that deconvolution-based PVC (RL and RVC) can markedly enhance radiomic reproducibility, they also underscore a broader methodological challenge in PET standardization. As highlighted by Cysouw et al. (30), the clinical translation of PVC remains controversial, although PVC improves quantitative accuracy in small lesions, its impact on diagnostic or prognostic performance across studies has been inconsistent. That meta-analysis revealed increased sensitivity but variable specificity after PVC, suggesting that the benefits depend strongly on lesion size, reconstruction protocol, and algorithm implementation rather than a universal improvement. Our findings extend this discussion to the radiomic domain: while PVC reduces variability and improves $ICC_2$ metrics, such improvements may partly stem from the controlled conditions of a single-center design and limited methodological diversity. Therefore, reproducibility alone cannot be equated with generalizability. Broader validation, incorporating multi-center data, motion-affected lesions, and newer deep-learning-based PVC frameworks, is essential to determine whether these gains translate into true robustness. Establishing standardized, reproducible PVC protocols across vendors and reconstruction pipelines remains a prerequisite for integrating PVC-corrected radiomic biomarkers into clinical decision-making.

Despite the promising reproducibility gains achieved through deconvolution-based PVC, several limitations should be acknowledged. First, this was a single-center study conducted using a single PET/CT system (GE Discovery 690) with uniform reconstruction parameters. While this homogeneity minimizes technical variability, it may artificially elevate reproducibility estimates and limit generalizability to multi-scanner or multi-vendor settings. Second, semi-automated segmentation, although refined by expert correction, introduces potential operator bias and reduces the independence of measurements, particularly in heterogeneous extranodal lesions. Third, only two classical iterative PVC algorithms (Richardson–Lucy and Reblurred Van Cittert) were evaluated; newer deep-learning-based PVC methods may further improve correction performance and should be compared systematically in future work. Fourth, we did not assess test–retest repeatability or multicenter harmonization, both of which are critical for validating radiomic biomarkers. Finally, as radiomic features were extracted under standardized conditions, further studies are needed to evaluate reproducibility under realistic clinical variability, including differences in acquisition protocols, motion effects, and reconstruction settings. Addressing these limitations through cross-center validation and inclusion of modern data-driven PVC frameworks will be essential for establishing reproducible and standardized PET radiomics pipelines.

## 5. Conclusion

In conclusion, partial volume correction substantially improves the reproducibility of $^{18}$F-FDG PET radiomic features in lymphoma. Larger and more homogeneous lesions demonstrated the highest stability, confirming that mitigating partial volume effects is essential for reliable radiomic quantification. Among all feature classes, First-Order and GLCM features showed the greatest robustness, supporting their suitability for prognostic and response-assessment models. While the strong reproducibility observed highlights the potential of PVC for standardizing PET radiomics, further multicenter and multi-method validation is required to ensure these improvements are consistent across scanners, protocols, and clinical settings.

**Acknowledgment**
The authors would like to thank Arzou Farahani Pour and Saeed Erfani Joo from Masih Daneshvari Hospital for their valuable support and assistance during this study.

Funding: Non

**Conflict of Interest**
The authors declare that there are no conflicts of interest regarding the publication of this manuscript.